# Blankets, Heat, and Why Free Energy Has Not Illuminated the Workings of the Brain

Donald Spector[1]
Daniel Graham[2]

[1] Department of Physics, Hobart and William Smith Colleges, Geneva, NY 14456, USA, spector@hws.edu, http://people.hws.edu/spector/
[2] Department of Psychological Sciences, Hobart and William Smith Colleges, Geneva, NY 14456, USA, graham@hws.edu, http://people.hws.edu/graham/

**ABSTRACT**

What can we hope to learn about brains from the free energy principle? In adopting the "primordial soup" physical model, Bruineberg et al. (2021) perpetuate the unsupported notion that the free energy principle has a meaningful physical – and neuronal – interpretation. We examine how minimization of free energy arises in physical contexts, and what this can and cannot tell us about brains.

**MAIN TEXT**

To determine the implications of applying free energy principles to the study of the brain, it is worth examining how free energy arises in physics in the first place, and then considering the implications for studies of the brain. We focus on two questions: What is the functional content of applying a free energy principle to the brain? If the free energy principle does work phenomenologically, can it tell us about the underlying workings of the brain?

Free energy arises in thermodynamics, the field that describes the bulk behavior of large systems. Statistical mechanics, in turn, is the field that derives thermodynamics from more fundamental principles. Via the ergodic hypothesis, statistical mechanics says that the bulk properties of a system (macrostates) can be found by ignoring the detailed dynamics of the intractably large number of microstates, and instead performing ensemble averages over the possible microstates (with equal likelihoods in isolated systems, which implies Boltzmann weightings at finite temperature). The bulk properties found by ensemble averages in statistical mechanics can alternatively be found thermodynamically, by minimizing the quantity known as the free energy.

The power of free energy is thus not that it is optimized at equilibrium; after all, there are non-thermodynamic optimization problems. It is that there is a language of macrovariables which can characterize a system, while the underlying microvariables evolve in a way functionally indistinguishable from randomly. But as invoked in the target article, the free energy principle does not point to any macro- or microvariables, which are needed for either a high- or low-level understanding of the workings of the brain.

If the free energy principle in the study of the brain is to be useful, we should hope that the process of deriving thermodynamics from statistical mechanics can be run in reverse: that establishing the efficacy of a free energy principle to describe the behavior and representational strategies of agents with a brain can reveal the fundamental dynamics of the brain. Even if we posit that there are microstates, all that is required for thermodynamics to arise is that the dynamics cause those microstates to be sampled over time with the correct weightings to allow the ensemble average to mimic the dynamics. Alas, this does not uniquely determine the underlying microstate dynamics.

Imagine thermodynamics had been invented before Newtonian mechanics. Could one deduce Newton's laws of motion from this formalism? The answer is no. For example, a gas at finite temperature can be modeled using kinetic theory or using a Metropolis algorithm. These provide different dynamical rules on microstates that produce the same thermodynamics; thermodynamics alone cannot reveal the fundamental dynamics. While broad results like entropy maximization arise in a general framework, to use statistical mechanics to obtain the thermodynamics of specific systems relies on knowledge about those systems extrinsic to thermodynamics, already obtained in other contexts.

Furthermore, even if one has posited micro-level dynamics for the brain, producing a thermodynamic language still requires identifying suitable macrostate variables. When tossing one million coins, if, instead of focusing on which particular coins are heads or tails (the microstates), we label states just by their total numbers of heads and tails, we can perform a free energy-style analysis to get the average behavior (and show fluctuations from this are negligible). This methodology hinges on choosing appropriate macrostate variables (e.g., the number of heads, not, say, the number of heads squared). Without a suitable analogous connection between microstates and macrostates, the promise of a free energy principle for the brain remains unfulfilled.

Of course, if the brain does achieve certain equilibrated behavioral states, one could by construction create a free energy function that said states minimize. Leaving aside the potential tautology of this philosophy, the question remains, what are those states? What macrovariables are static in equilibrium? Perhaps more importantly, how are they connected to the microstates of the brain? Should we focus on neuronal states or their interactions? Should we describe the brain in terms of synaptic events, spikes, spike timing, oscillations, local potentials, voxelwise patterns, or some combination of these? What microstates can be lumped together into useful macrostates, and by what rules?

Although the brain is complicated, accepting ignorance of its workings is untenable (imagine if thermodynamics itself had stopped with Carnot's generation and we never developed statistical mechanics and all that ensued). Still the free energy principle could be used to solve real-world problems with a set of well-understood affectors and effectors, i.e., in situations like neurorobotics where we do not necessarily want to model the brain but do want "intelligent" solutions to environmental challenges.

As we think about thermodynamics and brains, let us imagine how mysterious heat must have seemed at first. But heat, it turned out, was not a new form of energy, simply familiar forms of energy carried by degrees of freedom whose details were no longer being tracked. In studies of the brain, what plays the role of heat (or any other thermodynamic quantity), not literally, but as a seemingly distinct macro feature that embodies hidden micro behavior?

The free energy principle for brains is couched in the language of statistical mechanics but not justified by it. However, we would welcome attempts to work from brain microstates to a thermodynamic approach (and see what variables or principles are useful). Whatever the differences between the principles that prevail in brains and those relevant to physics, we still stand a better chance of understanding both the brain and behavior through the analogous study of principles in the brain as opposed to ensemble properties with unknown relationships to microstates and microstate dynamics. This is essentially the "inside-out" approach to systems neuroscience (Buzsáki, 2019). For example, what rules govern and how does the brain manage flexible, brain-wide communication flow on a neuronal network with short paths between essentially any populations of neurons (Graham, 2021)? If elucidating principles of brain function proves successful, we could interrogate the entire system of many elements and their interaction with the environment directly, and potentially dispense with blankets altogether.

**FUNDING STATEMENT**

This research received no specific grant from any funding agency, commercial or not-for-profit sectors.

**CONFLICTS OF INTEREST**

None